# Non-Local Realistic Theories
# and the Scope of the Bell Theorem

Federico Laudisa

*Department of Human Sciences, University of Milan-Bicocca,*

*Piazza dell'Ateneo nuovo 1, 20126 Milan, Italy*

e-mail: `federico.laudisa@unimib.it`

> We are told that no distinction is to be made between the state of a natural object and what I know about it, or perhaps better, what I can know about it if I go to some trouble. Actually – so they say – there is intrinsically only awareness, observation, measurement. If through them I have procured at a given moment the best knowledge of the state of the physical object that is possibly attainable in accord with natural laws, then I can turn aside as *meaningless* any further questioning about the "actual state", inasmuch I am convinced that no further observation can extend my knowledge of it. (Schrödinger 1935, p. 157)

> From those who made [the Copernican] revolution we learned that the world is more intelligible when we do not imagine ourselves to be at the centre of it. Does not quantum theory place observers…us…at the centre of the picture? There is indeed much talk of 'observables' in quantum theory books. And from some popular presentations the general public could get the impression that the very existence of the cosmos depends on our being here to observe the observables. I do not know that this is wrong. I am inclined to hope that we are indeed important. But I see no evidence that it is so in the success of contemporary quantum theory. (Bell (2004), p. 170)

## Abstract

According to a widespread view, the Bell theorem establishes the untenability of so-called 'local realism'. On the basis of this view, recent proposals by Leggett, Zeilinger and others have been developed according to which it can be proved that even some non-local realistic theories have to be ruled out. As a consequence, within this view the Bell theorem allows one to establish that no reasonable form of realism, be it local or non-local, can be made compatible with the (experimentally tested) predictions of quantum mechanics. In the present paper it is argued that the Bell theorem has demonstrably nothing to do with the 'realism' as defined by these authors and that, as a consequence, their conclusions about the foundational significance of the Bell theorem are unjustified.

# 1. Introduction

The question of how we should reshape the notion of physical reality after the advent of quantum mechanics continues to hold a central position in foundational debates, whereas the startling advances in experimental physics, and especially in quantum optics, seem to open up new ways of addressing the foundational issues in quantum mechanics. In particular, the scope of the Bell theorem and the exact nature of the constraints it prescribes for any consistent theory of quantum phenomena still remain crucial in most discussions even in very recent times.

In the April, 19 2007 issue of *Nature* an article was published (Gröblacher et al (2007)) in which a new experimental procedure was proposed for testing an inequality derived within a new class of theories, called *non-local realistic theories* (a class of theories originally introduced in Leggett (2003)). The authors could show that this inequality, which is derivable from the conditions of the above-mentioned non-local realistic theories, is at variance with the predictions of quantum mechanics: since – according to the authors – the Bell theorem shows that local realistic theories are incompatible with quantum mechanics, the conclusion of Gröblacher et al was that realism cannot be maintained even in a wide class of theories in which the locality requirement is relaxed.

As will be shown more in detail later, the whole enterprise depends crucially on the claim that the Bell theorem has within its premises *both* locality and a condition called 'realism', a condition which is often formulated, even recently, as the idea that physical systems are endowed with certain *pre-existing* properties, namely properties possessed by the systems prior and independently of any measurement interaction and that determine or may contribute to determine the measurement outcomes (Gröblacher S. et al (2007), p. 871). Although it has been clearly shown – from the original 1964 Bell paper right up to more recent instances (Maudlin (1996), Norsen (2007)) – that the Bell theorem does not include any 'realism' among its assumptions and that the non-locality established by the theorem holds for *any* theory that preserves quantum-mechanical predictions, be it 'realistic' or 'non-realistic', there seems to be a die-hard tendency to regard the Bell theorem as a result that does not establish non-locality but rather the impossibility of any objective (i.e. observer-independent in principle) account of the physical world, provided quantum mechanics is taken for granted. As a matter of fact, not only is the correct interpretation of the Bell theorem not fully acknowledged but also complex experimental settings are designed in important laboratories around the world, in order to test what appear as the implications of a clearly incorrect interpretation of the Bell theorem. Moreover, such ill-founded interpretations of one of the most relevant results for the whole field of the foundations of physics are disseminated – as the Gröblacher S. et al (2007) paper shows – in the most respected scientific journals.

All this suggests that a re-assessment of the question is still needed: the present paper is meant to show how the above-mentioned incorrect interpretation of the Bell theorem leads to carrying out the pursuit of implausible research programs on the foundations of quantum mechanics.

The plan of the paper is the following. In section 2 I will summarize the main claims contained in Gröblacher et al (2007) and concerning (**i**) the role of the so-called *local realism* in several versions of the Bell theorem (in fact the claim on this point represents views expressed implicitly or explicitly in many other more or less recent places disseminated in the literature, especially in the field of the quantum theory of information and computation); (**ii**) the prospects of investigating a new class of theories – called *non-local realistic theories* – as a consequence of the interpretation attached to the meaning of the Bell theorem according to (**i**). Sections 3 and 4 are devoted to a critical analysis of such claims: since the logically fundamental claim of these authors is that the Bell theorem is a consequence of assuming locality *and* realism, the aim of these sections is to show in what sense no *independent* 'realistic' assumption plays any role in establishing the conclusion of the Bell theorem, either in the *strict* (Bell (1964)) or in the *non-strict* correlation framework (Bell (1971), (1981))[1]. In doing this, two collateral but important points will be stressed. Firstly, not only was no 'realistic' assumption *de facto* required in the proof of the Bell theorem (either in deterministic or in stochastic settings), but also that a possible interpretation of 'realism' in terms of a *pre-existing property assumption* is inconsistent with quantum mechanics (Bell (1966)), no matter whether locality or non-locality are taken into account or not (Laudisa (1997)). Second, it will be stressed that Bell himself was perfectly clear about the irrelevance of any 'realistic' assumption for the derivation of his theorem (again, both in the deterministic and stochastic setting).

Finally in section 5, on the basis of the conclusions drawn in the preceding sections, I will argue against the relevance of assessing the compatibility with quantum mechanics of theories that are assumed to be non-local and yet realistic in the above mentioned ill-founded sense. I will also argue that endorsing such a sort of 'realism' leads first to the investigation of theories that are totally irrelevant from the viewpoint of the foundations of quantum mechanics, and secondly to overlook theories that are much more significant but which for very serious and structural reasons do not fall under the category of non-local realistic theories.

---

[1] I speak here of 'realism' in quotation marks precisely because the controversial matter is just what it takes to be 'realistic' toward the quantum mechanical description of physical systems.

## 2  On the role of local realism in the Bell theorem

The best place to start is with the summary of the situation as depicted by Gröblacher et al (2007):

> Bell's theorem proves that all hidden-variable theories *based on the joint assumption of locality and realism* are at variance with the predictions of quantum physics. Locality prohibits any influences between events in space-like separated regions, while *realism claims that all measurement outcomes depend on pre-existing properties of objects that are independent of the measurement*. The more refined versions of Bell's theorem By Clauser, Horne, Shimony and Holt and by Clauser and Horne start from the assumptions of local realism and result in inequalities for a set of statistical correlations (expectation values), which must be satisfied by all local-realistic hidden variable theories. The inequalities are violated by quantum mechanical predictions. [...] So far all experiments motivated by these theorems are in full agreement with quantum predictions [...] Therefore it is reasonable to consider the violation of local realism a well established fact. (p. 871, italics added)

In the authors' text, the expression 'Bell's theorem' without qualification refers to the original 1964 formulation by John S. Bell, in which the ideal experimental setting contemplated the emission of pairs of spin-1/2 particles prepared at the source in the spin singlet state. In this ideal setting the source state of the joint system prescribes a *strict* anticorrelation between the measurement outcomes in the two wings of the experimental setting, whereas the measurement outcomes were supposed to be associated with spacetime regions that are space-like separated (Bell (1964)). On the other hand, in the 'more refined versions' of Bell's theorem which the text refers to, the strict anticorrelation requirement is relaxed and this in turn paves the way toward an experimentally feasible test of the Bell inequality (Clauser, Horne, Shimony, Holt (1969), Bell (1971), Clauser, Horne (1974), Bell (1981)). In the Gröblacher et al (2007) approach, holding realism amounts to assuming the following:

> **REALISM$_{G\&AL}$** – The physical systems under scrutiny are endowed with *pre-existing* properties that (**i**) do not depend essentially on the measurement interactions the systems themselves may undergo, (**ii**) determine all the outcomes of possible measurements that can be performed on the physical systems.

The two points (**i**) and (**ii**) are reminiscent of the widespread terms 'Non-Contextuality' and 'Determinism', respectively. The point (**i**), in particular, seems to presuppose that, in order for a theory to be 'Realist $_{G\&AL}$', the pre-existing properties do not depend on the measurement interactions in that they are passively *revealed* by the measurements themselves.[2] Clearly, the subscript '**G&AL**' is meant to refer to the Gröblacher et al (2007) formulation of 'realism'.

As mentioned above, the idea that **REALISM**$_{G\&AL}$ is an independent condition under which – jointly with a locality condition – the Bell theorem can be proved is still a widespread idea, that can be found formulated in essentially the same terms in several (more or less recent) texts, although in most of these texts it is far from clear whether the authors really assume **REALISM**$_{G\&AL}$, namely both conditions (**i**) and (**ii**) of the above definition or just one of them. In his *Lectures on quantum theory* Chris J. Isham, for instance, claims that in the usual spin correlation framework of the Bell theorem "the central realist assumption we are testing is that each particle has a definite value at all times for any direction of spin" (Isham (1995), p. 215), and after the inequality has been derived we read (p. 216)

> It is important to emphasize that the only assumptions that have gone into proving [the inequality] are:
> 1. For each particle it is meaningful to talk about the actual values of the projection of the spin along any direction.
> 2. There is locality in the sense that the value of any physical quantity is not changed by altering the position of a remote piece of measuring equipment.

In their book on the foundations of quantum computations and information, after summarizing the lesson that is supposed to be drawn from the Bell theorem, Michael A. Nielsen and Isaac L. Chuang claim:

> What can we learn from Bell's inequality? For physicists, the most important lesson is that their deeply held commonsense intuitions about how the world works are wrong. The world is *not* locally realistic. Most physicists take the point of view that it is the assumption of realism which needs to be dropped from our worldview in quantum mechanics, although others have argued that the assumption of locality should be dropped instead. Regardless, Bell's inequality together with substantial experimental evidence now points to the conclusion that either or both of locality and realism must be dropped from our view of the world if we are to develop a good intuitive understanding of quantum mechanics. (Nielsen, Chuang (2000), p. 117)

Along similar lines, Asher Peres and Daniel Terno have argued that

---
[2] To be fair, Gröblacher et al (2007) are not entirely clear on this presupposition but I claim that my interpretation of their condition is the most reasonable if one wishes to preserve consistency with what they claim in their paper as their general conclusion.

Bell's theorem (1964) asserts that it is impossible to mimic quantum theory by introducing a set of objective *local* "hidden" variables. It follows that any classical imitation of QM is necessary nonlocal. However Bell's theorem does not imply the existence of any nonlocality in quantum theory itself." (Peres, Terno (2004), p. 104)

Cristopher Fuchs and Asher Peres have emphasized the same point by claiming that "John Bell formally showed that any *objective* theory giving experimental predictions identical to those of quantum theory would necessarily be nonlocal." (Fuchs, Peres (2002), p. 71)[3], whereas in a recent article devoted to the EPR argument in the so-called 'relational approach' to quantum mechanics, Smerlak and Rovelli formulate the issue in the following terms:

In the original 1935 article, the EPR argument was conceived as an attack against the description of measurements in Copenhagen quantum theory and a criticism of the idea that Copenhagen quantum mechanics could be a *complete* description of reality. Locality and a strong form of realism were given for granted by EPR and completeness was argued to be incompatible with quantum-mechanical predictions. With Bell's contribution, which showed that EPR correlations are incompatible with the existence of a hypothetical *complete* local realist theory, the argument has been mostly reinterpreted as a direct challenge to "local realism". [...] On the other hand, the Kochen-Specker theorem has questioned the very possibility of uncritically ascribing "properties" to a quantum system. From this perspective, the problem of locality moves to the background, replaced by a mounting critique of strongly objective notions of reality. Here we take this conceptual evolution to what appears to us to be its necessary arriving point: *the possibility of reading EPR-type experiments as a challenge to Einstein's strong realism, rather than locality*. (Smerlak, Rovelli 2007, p. 427, last italics added)[4]

On the basis of the above argument, then, a Bell-type inequality can be (**i**) derived from locality and **REALISM$_{G\&AL}$**, and (**ii**) shown to be contradicted by the statistical predictions of quantum mechanics. After a large number of experimental tests that confirm the latter[5], hence showing that the Bell inequality cannot be valid in the quantum domain, "it is reasonable to consider the violation of local realism a *well established fact*." (Gröblacher et al (2007), p. 871, italics added)

---

[3] See also Zukowski (2005), pp. 569-570. By 'objective' the authors in both the last quotations mean 'realistic' in the sense of our **REALISM$_{G\&AL}$** formulation. For examples of authors who correctly do *not* assume **REALISM$_{G\&AL}$** as an independent condition in referring to the Bell theorem, see for instance Squires (1986), pp. 83-91, and D'Espagnat (1995), pp. 142-3.
[4] Norsen (2007), pp. 312-314, interestingly traces a sort of true history of the claim according to which "local realism" is held to be the focus of the Bell theorem.
[5] References to the reports of such experiments can be found in Gröblacher et al (2007).

Since logically we would have an alternative between locality and **REALISM**<sub>G&AL</sub>, so that at least one of the two must be dropped, in the following, Gröblacher et al (2007) depict the prospects of any investigation taking seriously the above alternative:

> The logical conclusion one can draw from the violation of local realism is that at least one of its assumptions fails. Specifically, either locality or realism or both cannot provide a foundational basis for quantum theory. Each of the resulting possible positions has strong supporters and opponents in the scientific community. However, Bell's theorem is unbiased with respect to those views: on the basis of this theorem, one cannot, even in principle, favour one over the other. It is therefore important whether incompatibility theorems similar to Bell's can be found in which at least one of the these concepts is relaxed. Our work addresses a broad class of non-local hidden-variable theories that are based on a very plausible type of realism and that provide an explanation for all existing Bell-type experiments. Nevertheless we demonstrate, both in theory and experiment, their conflict with quantum predictions and observed measurement data. (Gröblacher et al (2007), pp. 871-2)

The new step then would be to investigate the viability of a class of theories that accept a weakening of the locality requirement while sticking to a 'very plausible type of realism'. The upshot of this line of research consists finally in proving, via new testable inequalities, that no matter how non-local our theory might be, we cannot adhere to any reasonable form of realism whatsoever if we agree to preserve the statistical predictions of quantum mechanics.

The specific features of this new class of non-local realistic theories were proposed for the first time in Leggett (2003). As a general premise, and in line with the above-mentioned quotations, Leggett claims that

> Bell's celebrated theorem states that, in a situation like that considered by Einstein *et al*., which involves the correlation of measurements on two spatially separated systems which have interacted in the past, no *local* hidden-variable theory (*or more generally, no objective local theory*) can predict experimental results identical to those given by standard quantum mechanics. (p. 1469, italics added)

Leggett proposes then introducing a class of *non-local* hidden-variable theories – namely a class of theories which, while retaining 'objectivity' (as will be seen later, it is the above-formulated **REALISM**<sub>G&AL</sub> condition), accept the incorporation of the possibility of non-local physical processes. The motivation for such a theoretical move is the following:

> In my view, the point of considering such theories is not so much that they are in themselves a particularly plausible picture of physical reality, but that by investigating their consequences one may attain a deeper insight into the nature of quantum-mechanical "weirdness" which Bell's theorem explores. *In particular I believe that the results of the present investigations provide quantitative backing for a point of view which I believe is by now certainly well accepted at the qualitative level, namely that the incompatibility of the predictions of objective local theories with those of quantum mechanics has relatively little to do with locality and much to do with objectivity.* (p. 1470, italics added)

The theories in the Leggett class are supposed to account for the results obtained in a general experimental framework, in which some polarization measurements are performed on pairs of photons emitted by atoms in a cascade process (Leggett (2003), p. 1471 ff). Since this framework encompasses, after the emission, a number of detection processes involving a pair of spatially separated detectors (let us call them **D**$_1$ and **D**$_2$), attention is focused as usual on correlations between the counts: clearly, the aim is to compare the predictions for a given function of such correlations as prescribed by quantum mechanics on the one hand and by (what Leggett assumes as) a general hidden-variable theory on the other.

The general conditions that the Leggett-type of theories are assumed to satisfy are the following (Leggett (2003), pp. 1473-4):

**L1**. Each pair of photons emitted in the cascade of a given single atom is characterized by a unique value of some set of hidden variables denoted by λ.

**L2**. In a given type of cascade process, the ensemble of pairs of emitted photon is determined by statistical distribution of the values of λ, characterized by a normalized distribution function ρ(λ). Such function is assumed to be *independent* of any parameter concerning polarizer settings (denoted in the sequel with **a** and **b**) and detection processes.

**L3**. If **A** and **B** denote respectively two variables that take the value + 1 (− 1) according to whether the detectors **D**$_1$ and **D**$_2$ register (do not register) the arrival of a photon, the value of **A** may depend not only on **a** and λ but also possibly on **b**, and similarly the value of **B** may depend not only on **b** and λ but also possibly on **a**.

Condition **L1** expresses the requirement (**i**) of REALISM$_{G\&AL}$, condition **L2** prevents the possibility of conspiratorial dependences between the source and any parameter involved in the spacetime regions where the polarizers and the detectors are located, whereas **L3** allows for possibly non-local influences of polarizer setting parameters on the outcomes[6]. Clearly this last condition, which according to the Leggett terminology characterizes the theories of the class as *crypto-nonlocal*, is

---
[6] It might be called 'Non-local determinism', since the actual outcomes are well determined by the pre-existing properties of the systems but possibly also in a non-local way.

where the theory is supposed to go beyond the class of theories ruled out by the Bell theorem (*according to the Leggett and followers' approach*)[7]. Jointly, **L1-L3** imply the following expression for the correlation to be measured P(**a**, **b**)

$$P(\mathbf{a}, \mathbf{b}) = \int_\Lambda \mathbf{A}(\mathbf{a}, \mathbf{b}, \lambda)\, \mathbf{B}(\mathbf{b}, \mathbf{a}, \lambda)\, \rho(\lambda)\, d\lambda$$

In addition to **L1-L3**, it is assumed that the local averages ⟨**A**⟩ and ⟨**B**⟩ agree with the relevant quantum mechanical predictions, which appears to be a rather natural 'consistency' condition on the Leggett class of hidden-variable theories (Leggett (2003), pp. 1476-9, Gröblacher et al (2007), p. 872). The last step is then the statement of an 'incompatibility' result consisting in the derivation, within crypto-nonlocal realistic hidden-variable theories, of an inequality that is violated by the corresponding quantum mechanical expressions (Leggett (2003), sect. 3).

Gröblacher et al (2007) elaborate a refinement of the Leggett framework by introducing a class of theories based on the following assumptions:

> (1) All measurement outcomes are determined by pre-existing properties of particles independent of the measurement (realism); (2) physical states are statistical mixtures of subensembles with definite polarizations, where (3) polarization is defined such that the expectation values taken for each subensemble obey Malus' law. (Gröblacher et al (2007), p. 872)

Since the theoretical framework is intended to be as general as to possibly allow some form of non-locality, Gröblacher et al (2007) assume that any individual measurement outcome for a polarization measurement along a fixed direction **u** is 'predetermined' not only by hidden variable λ (in addition to **u**) but also by some unspecified non-local parameter η, such that, if *A* denotes the measurement outcome, we have $A = A(\lambda, \mathbf{u}, \eta)$. Moreover, they introduce a probability distribution $\rho_\mathbf{u}(\lambda)$, by taking into account the possibility that particles with the same **u** might have different λ giving rise to subensembles of definite polarization. The final move is in two steps. First, on the basis of the above assumptions a further Leggett-type inequality is derived (Gröblacher et al (2007), p. 873). Second, a refined experimental setting is introduced that employs spontaneous parametric down conversion techniques, with the aid of which the Leggett-type inequality can be effectively tested against quantum mechanical predictions (Gröblacher et al (2007), pp. 874-5). The lesson to be learned, according to the last authors, is summarized in the conclusion of the article:

---

[7] As a matter of fact, the Leggett-type of theories are 'realistic' hidden variable theories that are assumed to be non-local by accepting outcome independence but dropping parameter independence (according to the Shimony revision of the terminology introduced in Jarrett (1984). I postpone to section 5 the discussion on how happily a **REALISM** assumption may coexist with parameter dependence. I wish also to stress that in presenting the Leggett framework I skip several technical details that, although deserving attention, are inessential to the present discussion.

> We have experimentally excluded a class of important non-local hidden-variable theories. In an attempt to model quantum correlations of entangled states, the theories under considerations assume realism, a source emitting classical mixtures of polarized particles (for which Malus' law is valid) and arbitrary non-local dependencies via the measurement devices. Besides their natural assumptions, the main appealing feature of these theories is that they allow us both to model perfect correlations of entangled states and to explain all existing Bell-type experiments. *We believe that the experimental exclusion of this particular class indicates that any non-local extension of local theory has to be highly counterintuitive. [...] Furthermore, one could consider the breakdown of other assumptions that are implicit in our reasoning leading to the inequality. They include Aristotelian logic, counterfactual definiteness, absence of actions into the past or a world that is not completely deterministic. We believe that our results lend strong support to the view that any future extension of quantum theory that is in agreement with experiments must abandon certain features of realistic descriptions*. (p. 875, italics added)

The aim of the subsequent sections is to show that these conclusions follow on from a totally misguided interpretation of the Bell theorem and that, as a consequence, cannot have the significance Leggett (2003) and Gröblacher et al (2007) (and all their followers) attach to them concerning the features of possible extensions – or simply consistent interpretations – of quantum theory.

## 3  REALISM in the strict anticorrelation framework

We have seen that according to both Leggett and Gröblacher et al, the heart of the Bell theorem is local realism and, as a matter of fact, all these authors refer explicitly to the celebrated article published by John S. Bell in 1964 when mentioning the Bell theorem. Curiously enough, the clearest and most useful starting point in order to see why their statements are wrong is exactly the opening of the 1964 Bell article (Bell (1964)). In the first pages, Bell summarizes the EPR-Bohm incompleteness argument in order to state unambiguously the premises from which his own non-locality theorem is to proceed. I will start first from the informal wording that Bell himself employs in stating the aim of his article, and I will proceed to a step-by-step formulation of the EPR-Bohm argument in order to show that a REALISM$_{G\&AL}$ condition is *derived* and not *assumed*. Finally I will quote the Bell summary of the situation, a summary that once again states clearly the derivative character of REALISM$_{G\&AL}$.

John S. Bell opens his article as follows:

The paradox[8] of Einstein, Podolsky and Rosen was advanced as an argument that quantum mechanics could not be a complete theory but should be supplemented by additional variables. These additional variables were to restore causality and locality. In this note that idea will be formulated mathematically and shown to be incompatible with the statistical predictions of quantum mechanics. *It is the requirement of locality, or more precisely that the result of a measurement on one system be unaffected by operations on a distant system with which it has interacted in the past, that creates the essential difficulty*. (Bell (1964), in Bell (2004), p. 14, italics added)

As is well known, the EPR-Bohm framework consists in a system $S_1+S_2$ of two spin-1/2 particles $S_1$ and $S_2$ prepared at time $t_0$ in the singlet spin state

$$\Psi = 1/\sqrt{2}\ (|1,+>_\mathbf{n} |2,->_\mathbf{n} - |1,->_\mathbf{n} |2,+>_\mathbf{n}),$$

where **n** is a generic spatial direction. We assume that spin measurements are performed on subsystems $S_1$ and $S_2$ at space-like separation. According to the standard rules of quantum mechanics, we know that

- **REDUCED STATES (RS)** if the state of $S_1+S_2$ is $\Psi$, then

  (reduced) state of $S_1 \to \rho(1,\Psi) = 1/2\ (\mathbf{P}_{|1,+>\mathbf{n}} + \mathbf{P}_{|1,->\mathbf{n}})$,

  (reduced) state of $S_2 \to \rho(2,\Psi) = 1/2\ (\mathbf{P}_{|2,+>\mathbf{n}} + \mathbf{P}_{|2,->\mathbf{n}})$,

  and, for any **n,**

  $\text{Prob}_{\rho(1,\Psi)}$ (spin $_\mathbf{n}$ of $S_1 = +1$) = $\text{Prob}_{\rho(1,\Psi)}$ (spin $_\mathbf{n}$ of $S_1 = -1$) = 1/2

  $\text{Prob}_{\rho(2,\Psi)}$ (spin $_\mathbf{n}$ of $S_2 = +1$) = $\text{Prob}_{\rho(2,\Psi)}$ (spin $_\mathbf{n}$ of $S_2 = -1$) = 1/2

- **PERFECT ANTICORRELATION (PAC)** If at time $t$ a spin measurement is performed on the subsystem $S_1$ in direction **n** and the outcome +1 [−1] is obtained, then a spin measurement on the subsystem $S_2$ in the same direction **n** at an immediately subsequent time $t' > t$ will have with certainty the outcome $-1$ [$+1$], namely

$$\text{Prob}_\Psi [(\text{spin}_\mathbf{n} \text{ of } S_1 = +1) | (\text{spin}_\mathbf{n} \text{ of } S_2 = -1)] =$$

$$\text{Prob}_\Psi [(\text{spin}_\mathbf{n} \text{ of } S_1 = -1) | (\text{spin}_\mathbf{n} \text{ of } S_2 = +1)] = 1.$$

---

[8] By the way: in his celebrated 1964 paper, Bell himself – and this is not the least important aspect in which we should appreciate his clear thinking – simply pays lip service to the use of the word 'paradox' in connection with the EPR arrangement; for in the second line of the first page of the paper he aptly stresses that we deal with an *argument*, namely a finite and ordered sequence of sentences whose validity we can assess by individuating clearly the premises and by checking whether the conclusion is a logical consequence of the premises themselves. As we will see briefly, the history of the misunderstandings in stating clearly which are the premises both of the EPR argument and of the derivation of the Bell-type inequalities is not over. And in the now classic 'Bertlmann's socks' paper (details below), Bell says: "And as if a child has asked: How come they always choose different colours when they *are* looked at? How does the second sock know what the first has done? Paradox indeed! But for the others, not for EPR. EPR did not use the word 'paradox'. They were with the man in the street in this business. For them these correlations simply showed that the quantum theorists had been hasty in dismissing the reality of the microscopic world." (Bell (2004), p. 143)

Let us suppose now that we perform at time $t_1 > t_0$ a spin measurement on $S_1$ in the direction **n** with outcome $+1$. Then, according to **PAC**, a spin measurement on $S_2$ in the same direction **n** at a time $t_2 > t_1$ will give with certainty the outcome $-1$. Let us suppose further to assume the following condition:

**PROPERTY-DEFINITENESS** - If, without interacting with a physical system $S$, we can predict with certainty - or with probability 1 - the value **q** of a physical quantity **Q** pertaining to $S$, then **q** represents an objective property of $S$ (denoted by [**q**]).

It is worth stressing that this condition amounts *not* to assuming the existence of objective properties, but rather to giving a *sufficient* condition for a property of a physical system to be 'objective'. In a nutshell:

(EPR-Bohm) **PROPERTY-DEFINITENESS** ≠ **REALISM** $_{G\&AL}$.

Then, at $t_2 > t_1$ [**spin**$_\mathbf{n}$ = $-1$] represents an objective property of $S_2$. But let us ask now: might the property [**spin**$_\mathbf{n}$ = $-1$] of $S_2$ have been somehow "created" by the spin measurement on $S_1$? The answer is clearly negative if we assume the following condition:

**LOCALITY** - No objective property of a physical system $S$ can be affected by operations performed on physical systems isolated from $S$.

Then, according to **LOCALITY**, the existence of the property [**spin**$_\mathbf{n}$ = $-1$] of $S_2$ can be inferred also relatively to a time $t'$ such that $t_0 > t' > t_1$. But at time $t'$ the state of $S_1+S_2$ is the singlet state $\Psi$. Therefore, according to (**RS**), the state of $S_2$ at time $t'$ is the reduced state

$$\rho(2,\Psi) = 1/2(\mathbf{P}_{|2,+>\mathbf{n}} + \mathbf{P}_{|2,->\mathbf{n}}),$$

a state that assigns to the property [**spin $_n$** = −1] of $S_2$ only the probability 1/2. Let us finally assume the following condition:

**COMPLETENESS** - Any objective property of a physical system *S* must be represented within the physical theory that describes *S*.

Then there exist objective properties of physical systems, such as [**spin $_n$** = −1] for $S_2$, that quantum mechanics is unable to represent: it follows that quantum mechanics is incomplete.

Let us ask now:

Is it assumed somewhere in the argument that properties such as [**spin $_n$** = − 1] pre-exist, namely they exist *independently*, *over and above* any spin measurement?

or

Does **PROPERTY-DEFINITENESS** imply only *by definition* that such properties as [**spin $_n$** = −1] pre-exist, namely they exist *independently*, *over and above* any spin measurement?

The answer to both questions is clearly negative! That such properties as [**spin $_n$** = −1] are to exist for $S_2$ *independently*, *over and above* any spin measurement is *not* assumed but rather is a *consequence* of the other assumptions of the argument. In fact,

- according to **PAC**

    Prob $_\Psi$ [(spin $_n$ of $S_1$ = +1) & (spin $_n$ of $S_2$ = −1)] = 1, for any **n**,

- according to **PROPERTY-DEFINITENESS**

    the outcomes of the spin measurements in the singlet state satisfy the condition of objective properties

- according to **LOCALITY**

    such outcomes have not been created by the distant measurement and **then** were 'already there'.

And here is the Bell summary:

Consider a pair of spin one-half particles created somehow in the singlet spin state and moving freely in opposite directions. Measurements can be made, say by Stern-Gerlach magnets, on selected components of the spins $\boldsymbol{\sigma}_1$ and $\boldsymbol{\sigma}_2$. If measurement of the component $\boldsymbol{\sigma}_1\bullet\mathbf{a}$, where **a** is some unit vector, yields the value +1 then, according to quantum mechanics, measurement of $\boldsymbol{\sigma}_2\bullet\mathbf{a}$ must yield the value -1 and vice versa. Now we make the hypothesis, and it seems one at least worth considering, that if the two measurements are made at places remote from one another the orientation of one magnet does not influence the result obtained with the other. Since we can predict in advance the result of measuring any chosen component of $\boldsymbol{\sigma}_2$, by previously measuring the same component of $\boldsymbol{\sigma}_1$, *it follows* that the result of any such measurement must actually be predetermined. Since the initial quantum mechanical wave function does not determine the result of an individual measurement, this predetrmination implies the possibility of a more complete specification of the state. (Bell (1964), in Bell (2004), pp. 14-15, italics added)

As should be clear from a fair reading of the Bell original article, the Bell theorem starts exactly from the alternative established by the EPR-Bohm argument – namely, locality and completeness cannot stand together – and goes for the proof that, *whatever form the completability of quantum mechanics might assume*, the resulting theory cannot preserve the statistical predictions of quantum mechanics and be local at the same time: this means that neither a pre-existing-property assumption (or **Objectivity** or **Classicality** or whatever synonymous one likes to choose) nor a determinism assumption are assumed in the derivation of the original Bell inequality. Therefore all claims – Leggett (2003) and Gröblacher et al (2007) included – to the effect that the Bell theorem in the 1964 setting concerns 'local *realism*' are completely ill-founded. As Bell himself (*vox clamantis in deserto*) clearly stresses:

It is important to note that to the limited degree to which *determinism* plays a role in the EPR argument, it is *not assumed* but *inferred*. What is held sacred is the principle of 'local causality' - or 'no action at a distance'. [...] It is remarkably difficult to get this point across, that determinism is not a *presupposition* of the analysis." […] My own first paper on this subject [Bell refers here to his 1964 paper] starts with a summary of the EPR paper *from locality to deterministic hidden variables*. But the commentators have almost universally reported that it begins with deterministic hidden variables. (Bell 1981, in Bell (2004), pp. 143, 157 footnote 10, italics in the original)

To sum up, the true logic of the argument is the following ('PP' stands for 'Pre-existing properties'):

      **1**. QM $\wedge$ LOC $\to$ PP           [EPR-Bohm Argument]

      **2**. PP $\to \neg$ QM                 [Bell Theorem][9]

      **3**. QM                         [Assumption]

---

[9] For an extremely easy and compact formulation of the Bell theorem for the strict correlation case, see Dürr, Goldstein, Zanghì (2004).

| | |
|---|---|
| **4**. QM → ¬ PP | [**2**, **3** *Modus tollens*] |
| **5**. ¬ PP | [**3**, **4** *Modus ponens*] |
| **6**. ¬ PP → ¬ (QM ∧ LOC) | [**1**, **5** *Modus tollens*] |
| **7**. ¬ (QM ∧ LOC) | [**5**, **6** *Modus ponens*] |

∴ ¬ LOC

But there is more to this question. That **REALISM**$_{G\&AL}$ cannot be a reasonable *independent* assumption of any allegedly 'objective' theory of quantum phenomena can be argued on the basis of what was already clearly demonstrated by Bell himself in the article that *preceded* the Bell theorem article, although as is well known, it was published *after* it (Bell (1966)). In this fundamental article Bell showed that all existing no-hidden variable theories proofs (Gleason, Jauch-Piron, Kochen-Specker and an additional proof provided by Bell himself as a simplified version of the Kochen-Specker theorem) required assumptions that it was not reasonable to require from any hypothetical completion of quantum theory[10]:

> It will be urged that these analyses [i.e. the above mentioned proofs] leave the real question untouched. In fact it will be seen that these demonstrations require from the hypothetical dispersion free states, not only that appropriate ensembles thereof should have all measurable properties of quantum mechanical states, *but certain other properties as well*. These additional demands appear reasonable when results of measurement are loosely identified with properties of isolated systems. They are seen to be quite unreasonable when one remembers with Bohr 'the impossibility of any sharp distinction between the behaviour of atomic objects and the interaction with the measuring instruments which serve to define the conditions under which the phenomena appear'. (Bell (1966), in Bell (2004), pp. 1-2, italics added)

If **REALISM**$_{G\&AL}$ were an *independent* assumption of any hidden variable theory, Gleason-Bell-Kochen & Specker would have already proved their incompatibility with quantum mechanics *needless of any locality requirement*. But, as Bell showed, there is little significance in testing against quantum theory a theory (be it local or non-local) that is supposed to satisfy a condition that we already know quantum mechanics cannot possibly and reasonably satisfy[11]. The same point occurs interestingly in another Zeilinger article, a short essay published on *Nature* in 2005 and entitled *The message of the quantum* (Zeilinger (2005)). Here, the author, after claiming once again that "John Bell showed that quantum prediction for entanglement are in conflict with local realism", argues:

> Most physicists view the experimental confirmation of the quantum predictions as evidence for nonlocality. But I think that the concept of reality itself is at stake, a view that is supported by the Kochen-Specker paradox. This observes that even for single particles it is not always possible to assign definite measurement outcomes, independently of and prior to the selection of specific apparatus in the specific experiment. (p. 743)

---

[10] Bell also mentions an especially restrictive assumption of the von Neumann theorem, an ssumption which makes the von Neumann formulation much stronger with respect to the non-contextual formulations given by Gleason, Jauch-Piron, Bell and Kochen-Specker, and hence even less plausible (for a detailed analysis of the von Neumann 'no-go' theorem see Giuntini, Laudisa (2001).

Curious argument indeed! Zeilinger *first* uses the Kochen-Specker theorem in order to support the idea that it is not possible to ascribe pre-existing properties even to single systems (making the unwarranted assumption that the only logically consistent way of defining realism is in terms of pre-existing properties) and *then* he takes seriously the project of experimentally testing the incompatibility between quantum mechanics and a theory that is non-local *and* realistic in the sense of what the Kochen-Specker theorem *prohibits* one from assuming. Why then worry about the confirmation of quantum prediction in laboratories and not be content with the Kochen-Specker theorem itself?[12]

## 4 REALISM in the non-strict anticorrelation framework

Although the Bell 1964 article is always cited as the *locus classicus* for the non-locality theorem, its formulation is not fully general. In fact, the ideal setting outlined in the paper crucially relies on strict anticorrelation, whereas subsequent investigations have explored the possibility of dropping it (also in view of an experimental realization of the setting itself). Very much in the spirit that Gröblacher et al would have voiced in 2007, it has been claimed (Žukowski (2005)) that also in these more general frameworks (in which a class of so-called stochastic hidden-variable theories was introduced) a **REALISM**$_{G\&AL}$ condition was among the assumptions that led to the derivation of a more general inequality, an inequality that can be shown to be violated by the corresponding statistical predictions of quantum mechanics.

---

[11] This point had been stressed in Laudisa (1997), where a discussion of the relationship between the Bell 1966 and the Bell 1964 articles – from the point of view of the consistency of any hidden variable approach to quantum theory – can be found. Bell recalls this point still in the opening page of his 1964 paper: "There have been attempts to show that even without such a separability or locality requirement no 'hidden variable' interpretation of quantum mechanics is possible. These attempts have been examined elsewhere and found wanting [Bell refers here to his 1966 article]" (Bell (1964), in Bell (2004), p. 14). Most recently, the same charge has been clearly stated and motivated in Norsen (2007), pp. 317-8, where my condition (**i**) in the formulation of REALISM$_{G\&AL}$ is called 'naive realism'.

[12] For further critical remarks on this Zeilinger article see Daumer, Dürr, Goldstein, Maudlin, Tumulka, Zanghì (2006).

In the stochastic hidden-variable theories' framework (originally introduced in Bell (1971) and Clauser, Horne (1974)), a typical joint system $S_1+S_2$ is prepared at a source – very much like the polarization process situation introduced in section 2 and concerning the Leggett approach – so that a 'completion' parameter $\lambda$ is associated with the single and joint detection counts. Suppose we denote by **a** and **b** respectively the setting parameters concerning two detectors, located at space-like separation and devised to register the arrival of $S_1$ and $S_2$ respectively. The model then is assumed to satisfy the following conditions:

- BCH1. $\lambda$ is distributed according to a function $\rho(\lambda)$ that does *not* depend either on **a** or on **b**.
- BCH2. The parameter $\lambda$ prescribes single and joint detection *probability*.
- BCH3. *Locality* holds, namely the $\lambda$-induced probability for the measurement outcomes for $S_1$ and $S_2$ separately is such that (**i**) the detection probability for $S_1$ depends only on $\lambda$ and **a**, (**ii**) the detection probability for $S_2$ depends only on $\lambda$ and **b**, (**iii**) and the joint detection probability is simply the product of the detection probability for $S_1$ and the detection probability for $S_2$.

According to the view presupposed in Leggett (2003), Žukowski (2005) and Gröblacher et al (2007), these stochastic hidden variable theories include a form of **REALISM**$_{G\&AL}$ among their assumptions. Žukowski (2005), for instance, focuses on the Bell discussion of the motivations underlying this framework (Bell (1981)) and reformulates his assumptions "in today's wording":

*Realism.* To put it short: results of unperformed measurements have certain, unknown but fixed, values. In Bell wording this is equivalent to the hypothesis of the existence of hidden variables.

*Locality.* "The direct cause (and effects) of events are near by, and even indirect causes (and effects) are no further away than permitted by the velocity of light" (p. 239), in short, events and actions in Alice's lab cannot influence directly simultaneous events in Bob's lab and his acts, etc.

"*Free Will*". The settings of local apparata are independent of the hidden variables (which determine the local results) and can be changed without changing the distribution of local hidden variables (p. 154). In short, Alice and Bob have a free will to fix the local apparatus settings, or more mildly, one can always have a stochastic process that governs the local choices of the settings, which is statistically independent from other processes in the experiment (especially those fixing the hidden variables). (Žukowski (2005), pp. 569-570)

On the basis of this set of assumptions, Žukowski argues that the Bell theorem (in the general stochastic formulation) has been 'overinterpreted' (Žukowski (2005), p. 571). The logical structure of the Žukowski reconstruction is the following. If we denote with R, LOC and FW the above assumptions of Realism, Locality and Free Will, respectively, with BI the Bell Inequalities and with QM the assumption of the validity of the statistical predictions of quantum mechanics, we have

| | | |
|---|---|---|
| **0.** | R, LOC, FW | [Assumptions] |
| **1.** | R ∧ LOC ∧ FW → BI | [Bell Theorem (in the Žukowski interpretation)] |
| **2.** | QM | [Assumption] |
| **3.** | QM → ¬ BI | [Experimental fact] |
| **4.** | ¬ BI | [**2**, **3** *Modus ponens*] |
| **5.** | ¬ BI → ¬ R ∨ ¬ LOC ∨ ¬ FW | [**1**, **4** *Modus tollens*] |
| **6**. | ¬ BI → ¬ R ∨ ¬ LOC | [FW is an assumption] |
| ∴ | ¬ R ∨ ¬ LOC | |

That is, since FW seems out of question, therefore the dilemma ¬ R ∨ ¬ LOC remains. It is at this point that the 'overinterpretation' paradox comes in. Since, according to the derivation above, we are left with the alternative ¬ R ∨ ¬ LOC, Žukowski sees no compelling justification for dropping LOC rather than R. In Žukowski's view the 'overinterpretation' of the Bell theorem would be exactly the 'automatic' move from ¬ R ∨ ¬ LOC to ¬ LOC, a move that in logical terms is not necessary .[13]

> Here comes another paradox: the consequences of Bell's theorem as they are now most frequently presented to the entire physics community. [...] There is nothing in the quantum formalism that would necessarily imply non-locality" (Žukowski (2005), pp. 571-572).

---

[13] According to logic alone, of course, from ¬ R ∨ ¬ LOC you can derive *both* ¬ R and ¬ LOC if the negation of neither has been derived earlier. The real point is the choice of the premises: as will be shown shortly, R need not be an independent premise and hence the derivation does not yield ¬ R ∨ ¬ LOC but rather ¬ LOC.

The conclusion is obvious: why not drop R instead, keeping locality together with quantum mechanics?[14]

A fair reading of the Bell argument in his 1981 article shows that the above conclusion by Žukowski is totally unwarranted. There Bell envisages the possibility of introducing an EPR-Bohm set-up in very general terms, in which we are interested in the joint probability distribution

$$P(A, B \mid a, b),$$

where each A and B may be a 'yes' or a 'no' and *a* and *b* stand respectively for two possible adjustable parameters (with the obvious interpretation). No mention of what sort of systems are involved need be made, and once some sort of BCH1-BCH3 conditions are assumed, it is easy to show the derivation of a CHSH inequality. Before deriving the inequality, in order to make clear what the real assumptions in the argument are and how general the presentation is intended to be, Bell explicitly states:

> Despite my insistence that the determinism was *inferred rather than assumed* [N.d.R. a new hint at the frequent misunderstandings of this *inference* in the original EPR and in his 1964 paper], you might still suspect somehow that it is a preoccupation with determinism that creates the problem. Note well that *the following argument makes no mention whatever of determinism* […] Finally you might suspect that the very notion of particle, and particle orbit has somehow led us astray […] So the following argument will not mention particles, nor indeed fields, nor any particular picture of what goes on at the microscopic level. Nor will it involve any use of the words 'quantum mechanical system", which can have an unfortunate effect on the discussion. *The difficulty is not created by any such picture or any such terminology. It is created by the predictions about the correlations in the visible outputs of certain conceivable experimental set-ups*." (Bell (1981), in Bell (2004), p. 150, italics added)

The conclusion to be drawn from the Bell discussion is twofold: first, *nowhere* in the Bell-CH arguments does the *Realism* assumption play any role; second, interpretations such as Žukowski's are affected by the prejudice that a metaphysical and totemic notion of Microphysical Reality was what Bell preoccupied himself with. Moreover, in order to explain in what sense just locality is the focus of the argument, Bell (1981) draws an example from ordinary life. Suppose we find a correlation between the rate of heart attacks *h* in two different and distant towns called A and B, namely

$$p(h_A, h_B) \neq p(h_A)\, p(h_B)$$

---

[14] This echoes very closely the view expressed in a passage of the Leggett (2003) article quoted above: "I believe that the results of the present investigations provide quantitative backing for a point of view which I believe is by now certainly well accepted at the qualitative level, namely that the incompatibility of the predictions of objective local theories with those of quantum mechanics has relatively to do with locality and much to do with objectivity." (Leggett (2003), p. 1470)

where clearly $h_A$ and $h_B$ denote respectively the rate of heart attacks in A and in B. Since A and B are supposed to be so far away from each other that it is not imaginable at the outset that there is some direct influence at work, a sound scientific attitude would lead us first – Bell claims – to make the hypothesis that there are some factors that contribute *locally* to the apparently correlated rates. Let us call these collective factors $L_A$ and $L_B$. According to this locality assumption, it will be then reasonable to assume

$$p(h_A, h_B \mid L_A, L_B, \lambda) = p(h_A \mid L_A, \lambda) \, p(h_B \mid L_B, \lambda)$$

where $\lambda$ denotes collectively any other relevant variables.

The attitude toward the justification of the locality condition in terms of a similar 'factorizability' in the derivation of the Bell inequality for stochastic hidden variables models is essentially the same: "let us suppose that the correlations in the EPR experiment are likewise «locally explicable» (Bell (1981), in Bell (2004), p. 152). Namely, the core of the argument lies in stating what preventing any action-at-a-distance amounts to, whatever the factors at A and B might be. The above assumption need not be grounded on the additional assumption that there are some *pre-existing properties* in the common past of the relevant events at A and B that enhance the correlation[15]. Such assumption would be certainly sufficient for the assumption of existence of local factors, but *not necessary*. Namely, it is true that the assumption of pre-existing properties for the two systems at the source might well imply locality, but the assumption that only local operations and influences can contribute to fix the single detection probabilities need *not* follow from pre-existing properties, and rightly so: as we stressed in the previous section, assuming pre-existing properties in a model that is to be tested against quantum mechanics, when quantum mechanics itself prevents us from allowing pre-existing properties when describing physical interactions in its own proper terms, would deprive the model under scrutiny of any foundational significance.

If the whole point of the Bell-CH arguments is then in fact to show that the correlations between the results A and B are not locally explicable, no matter what the relation is between A and B on one side and some allegedly 'objective' or 'pre-existing' properties corresponding to them on the other, we can safely say that also in the more general (no perfect correlation) case, *there is no 'local realism' at stake*. Logically, the argument then proceeds as follows

| | | |
|---|---|---|
| **0.** | LOC, FW | [Assumptions] |
| **1.** | LOC $\wedge$ FW $\rightarrow$ BI | [Bell theorem] |
| **2.** | QM $\rightarrow \neg$ BI | [Experimental fact] |

---

[15] A similar point, although relative to the derivation of the CHSH inequality in Clauser, Horne, Shimony, Holt (1969), has been raised by Norsen (2007), p. 319.

| | 3. | QM | [Assumption] |
|---|---|---|---|
| | 4. | $\neg$ BI | [**2**, **3** *Modus ponens*] |
| | 5. | $\neg$ BI $\rightarrow \neg$ LOC $\vee \neg$ FW | [**1**, **4** *Modus tollens*] |

$$\therefore \quad \neg \text{ LOC}$$

## 5  On the significance of non-local REALISTIC theories

We are now in a position to assess the prospects and the significance of the investigations on the new class of non-local **REALISTIC**$_{\text{G\&AL}}$ hidden-variable theories, the class for which Leggett (2003) claims the proof of a new 'incompatibility' theorem, the latest in a long series of 'no-go theorems' about quantum mechanics. But let me sum up first the conclusions established in the preceding sections.

(**1**)    The condition that we have referred to as **REALISM**$_{\text{G\&AL}}$ is not a reasonable condition to require from any meaningful hidden variable theory, since it would make the confrontation between such theory and quantum mechanics totally uninteresting. Should **REALISM**$_{\text{G\&AL}}$ be required, quantum mechanics would be in outright contradiction with the hidden variable theory, no matter whether any statistical predictions are taken into account or whether any experiment is carried out, but ruling out such a vacuous hidden variable theory would not teach us any useful lesson about the foundations of quantum mechanics. A clear formulation of this fact is already contained in the two groundbreaking – but still misunderstood! – articles by John S. Bell published in 1964 and 1966.

(**2**)    Even if *ex absurdo* we suppose that **REALISM**$_{\text{G\&AL}}$ is a reasonable requirement, it can be shown (and this is to be credited again to John S. Bell) that such requirement plays no *fundamental* role in the Bell theorem, either in its strict correlation version or in its non-strict correlation version. If this is the case, the meaning of the Bell theorem lies not in its casting light on how far we should go in renouncing our cherished 'realistic' view of the microworld if we want to maintain the statistical and empirical content of quantum theory, but rather in demonstrating once and for all that any theory (*be it endowed with 'hidden variables' or not*) that is to save the agreement with the statistical predictions of quantum mechanics must be non-local. What non-locality exactly will entail in a specific theory will depend on the particular structure and conceptual resources of the theory, but two features will have to be part of any such theory: first, the theory will have to take into account non-locality as a basic property at least of the world of microscopic systems and, second, if the theory is introduced as endowed with a clear ontology – namely, endowed with some clear indications as to the nature and structure of that basic spacetime inventory of the world that the theory is supposed to be about – the ontology itself could not possibly be simply 'realistic' in the unreasonable sense of **REALISM**$_{\text{G\&AL}}$.

Under **(1)** and **(2)**, the research program of non-local **REALISTIC**$_{\text{G\&AL}}$ hidden-variable theories can then have no foundational significance, in that we can hardly learn anything from the attempt of establishing the compatibility or incompatibility of quantum mechanics with a class of theories satisfying such unreasonable assumptions. As a consequence, if **REALISM**$_{\text{G\&AL}}$ is not a reasonable assumption to make for any significant alternative theory that is to be tested against quantum mechanics, this will hold both for *local* **REALISTIC**$_{\text{G\&AL}}$ hidden-variable theories and for *non-local* **REALISTIC**$_{\text{G\&AL}}$ hidden-variable theories.

But let us go further and take a closer look at the motivations that Leggett himself discusses in support of the assumptions satisfied by his non-local **REALISTIC**$_{\text{G\&AL}}$ hidden-variable theories. As we recalled earlier (section 2), the theories in the Leggett class are supposed to account for the results obtained in a general experimental framework, in which some polarization measurements are performed on pairs of photons emitted by atoms in a cascade process. The Leggett-type of theories are such that each pair of photons emitted in the cascade of a given single atom is characterized by a unique value of some set of hidden variables: this value, denoted by λ, is characterized by a normalized distribution function ρ(λ) and is assumed to be *independent* of any parameter concerning polarizer settings (denoted in the sequel with **a** and **b**) and detection processes. Finally, the Leggett-type of theories may display some non-locality: if **A** and **B** denote respectively two variables that take the value + 1 (− 1) according as the detectors **D**$_1$ and **D**$_2$ register (do not register) the arrival of a photon, the value of **A** may depend not only on **a** and λ but also possibly on **b**, and similarly the value of **B** may depend not only on **b** and λ but also possibly on **a** (Leggett (2003), p. 1471 ff).

The non-locality assumption deserves a discussion on its own. Leggett introduces his 'new' class of hidden variable theories by subtraction, so to say. In fact, he first introduces *local* **REALISTIC**$_{\text{G\&AL}}$ hidden-variable theories, namely theories that satisfy **L1** and **L2** and for which the following equalities hold

**Outcome Independence**

$$A(\mathbf{a}, \mathbf{b}, \lambda, B) = A(\mathbf{a}, \mathbf{b}, \lambda),$$

$$B(\mathbf{b}, \mathbf{a}, \lambda, A) = B(\mathbf{b}, \mathbf{a}, \lambda)$$

**Setting Independence**

$$A(\mathbf{a}, \mathbf{b}, \lambda, B) = A(\mathbf{a}, \lambda, B)$$

$$B(\mathbf{b}, \mathbf{a}, \lambda, A) = B(\mathbf{b}, \lambda, A)$$

Clearly, under **Outcome Independence** (**SI**) and **Setting Independence** (**OI**), the expression for the correlation P(**a**, **b**) to be measured becomes

$$P(\mathbf{a}, \mathbf{b}) = \int_\Lambda A(\mathbf{a}, \mathbf{b}, \lambda, B)\, B(\mathbf{b}, \mathbf{a}, \lambda, A)\, \rho(\lambda)\, d\lambda = \int_\Lambda A(\mathbf{a}, \lambda)\, B(\mathbf{b}, \lambda,)\, \rho(\lambda)\, d\lambda,$$

namely, the usual expression for the locality assumption in stochastic hidden variable models. The Leggett-type of theories, in addition to **L1** and **L2**, are assumed to satisfy **Outcome Independence** but in general fail to satisfy **Setting Independence**. That is, the Leggett framework inherits the interpretation of locality as a conjunction of **SI** and **OI** (Jarrett (1984))[16] and then proposes investigating the compatibility with quantum mechanics of a theory which – although 'realistic' – is non-local in the sense of being possibly setting-*dependent* (Leggett (2003), p. 1474). Curiously enough, however, the assumption of **Outcome Independence** is motivated by the following statement:

> I shall rather arbitrarily assert assumption (4) (outcome independence), The reason for doing it is not so much that it is particularly "natural" [...] but it is a purely practical one; if one relaxes (4) it appears rather unlikely (though I have no rigorous proof) that one can prove anything useful at all, and in particular it appears very likely that one can reproduce the quantum-mechanical results for an arbitrary experiment. (Leggett (2003), p. 1475)

In fact, it *can* be proved that quantum mechanics violates **Outcome Independence**[17]. Then, although it is conceivable that a model intended to be 'strongly' non-local is formulated to be – as it were – 'more non-local' than quantum mechanics itself is supposed to be (namely by dropping

---

[16] Personally I do not find the 'peaceful coexistence' strategy (for instance, Shimony (1984)), relying on the Jarrett distinction, either convincing or illuminating on the issue of how quantum mechanical non-locality is supposed to coexist with special relativity (see for instance Maudlin (2002$^2$), pp. 93-98). Here I presuppose it only the sake of discussion.

[17] See for instance Butterfield, Fleming, Ghirardi, Grassi (1993).

**Setting Independence**, which quantum mechanics *satisfies* [18]), on the other hand, if the model is assumed to satisfy **Outcome Independence**, the incompatibility between the correlations prescribed by the model and the quantum correlations might always be ascribed to the circumstance that quantum mechanics does *not* satisfy **Outcome Independence**. This in turn would not allow us to conclude anything about the failure or the survival of REALISM$_{G\&AL}$, which presumably was the aim of the whole model.

Moreover, there *is* a further 'hidden variable' model which satisfies **Outcome Independence** but not **Setting Independence** – namely *Bohmian mechanics*[19], whose consistency directly refutes the claims of Leggett and followers. In fact, Bohmian mechanics satisfies REALISM$_{G\&AL}$ and nevertheless provides a perfectly consistent account for all phenomena that quantum mechanics is able to treat unambiguously (Goldstein (2001)), providing in addition a clear and law-governed ontology of particles evolving in spacetime. How can it be? On the one hand, the measurement outcomes in Bohmian mechanics are determined by pre-existing, measurement-independent properties of the measured system, namely the precise positions of the particles in the system and, of course, the wavefunction[20]. On the other hand,

> in Bohmian mechanics the random variables $Z_E$ giving the results of experiments $E$ depend, of course, on the experiment, and there is no reason that this should not be the case when the experiments under consideration happen to be associated with the same operator. Thus with any self-adjoint operator $A$, Bohmian mechanics naturally may associate many different random variables $Z_E$, one for each different experiment $E \to A$ associated with $A$. A crucial point here is that the map $E \to A$ is many-to-one" (Dürr, Goldstein, Zanghì (2004), p. 1040).

So, Leggett, Gröblacher and the others fail to appreciate that the consistency of Bohmian mechanics is a direct refutation of their approach since they appear to assume that the preexisting properties that determine the outcome must somehow mathematically resemble the eigenstates of Hermitian operators. But that very specific claim is surely no part of "realism". One needs to note that standard quantum theory associates physically different experimental set-ups with the same Hermitian operator ("observable"). But it is no part of "realism" to demand that physically different set-ups be treated alike: the way that the pre-existent positions determine the outcome of an experiment may of course depend on just how the experiment is set up.

---

[18] See again Butterfield, Fleming, Ghirardi, Grassi (1993).
[19] See for instance Dürr, Goldstein, Zanghì (1996), Goldstein (2001).
[20] See for instance the discussion of exactly how the pre-existent locations of particle determine the outcomes of "spin measurements" in Albert (1992).

However, Gröblacher et al (2007) very briefly mention Bohmian mechanics with a highly dismissive attitude. But there are here at least three additional points that deserve attention. First, Gröblacher et al (2007) refer to Bohmian mechanics in its old-fashioned formulation with quantum potential, but they do not seem to be even aware that the contemporary formulation of the theory, known as *Bohmian mechanics*, can do totally without any quantum potential[21]. Second, they underrate the circumstance that even in its quantum potential formulation Bohmian mechanics is a counterexample to their general claims. Third, they overlook the fact that Bohmian mechanics has a clear explanation of why the theory satisfies **Outcome Independence** and fails to satisfy **Setting Independence** (see again for instance Maudlin ($2002^2$), p. 94) whereas quantum mechanics has no such explanation as to why it satisfies **Setting Independence** and violates **Outcome Independence**, and the Leggett-type non-local realistic hidden variable theory is unable to account in an intelligible way for the validity of **Outcome Independence** and the failure of **Setting Independence**. Presumably it is the biased claim that, in order for a theory to be objective in some meaningful sense, a **REALISM$_{G\&AL}$** condition must be assumed, that motivates why neither Leggett (2003) nor Gröblacher et al (2007) (nor all other defenders of **REALISM$_{G\&AL}$** as a reasonable assumption for hidden variable theories) take into account one of the few *seriously* objective (namely observer-free) interpretations that are both consistent and alternative to quantum mechanics in a viable sense.

**Conclusions**

As I have tried to show in the preceding sections, there is a strong prejudice surrounding the foundational meaning of the Bell theorem, a prejudice that seems to survive intact through the years, in spite of the clear statements to the contrary repeatedly expressed – to begin with – by the inventor of the theorem itself, namely John S. Bell. This prejudice not only survives but in the last years has become even stronger, supported as it is by an emphasis on quantum computation that tends to dissolve all deep conceptual problems of standard quantum theory into a new information-theoretic orthodoxy[22]. According to this prejudice, the core of the Bell theorem concerns a philosophical notion – realism – and proves that such notion is untenable on *physical* grounds, namely holding to it implies quantitative predictions that are contradicted by the quantum predictions. This interpretation sounds very much like the ultimate death sentence for realism, and such a sentence seems hard to resist since it is *physics* which pronounces it, namely the queen of the hard sciences. As argued above, this approach overlooks the circumstance that, in order to assess the implications of a theorem, we have to be clear about the conditions under which the theorem can be proved, and one need not be a physicist to acknowledge it. What logical soundness and physical reasonableness suggest (sections 3-5) is that the role of Bell's theorem is not to set constraints on how 'realist' we are allowed to be about quantum systems but rather, much more interestingly, to characterize a structural property of any theory that aims to cover the domain of validity covered so far by quantum mechanics, namely non-locality. As a consequence, whether a theory aiming to supersede quantum theory will be 'realist', 'non-realist', 'half-realist' or 'one-third realist', this will concern the further conceptual and formal resources of that theory and not at all the Bell theorem.

---

[21] For a survey, see Dürr, Goldstein, Zanghì (1996) and for an illuminating analysis of the superiority of the recent formulation over the old one, see Goldstein (1996), pp. 156-160.
[22] For an interesting and recent assessment of this tendency see Hagar (2007).

**Acknowledgements.** I would like to thank Detlef Dürr, Shelly Goldstein, Nino Zanghì and the anonymous referees for extremely helpful remarks on earlier versions of the present paper. Linguistic help from Stephen Bush is also gratefully acknowledged.

# References

1. Albert D. (1992), *Quantum Mechanics and Experience*, Harvard University Press, Cambridge, MA.
2. Bell J.S. (1964), "On the Einstein-Podolsky-Rosen paradox" *Physics* **1**, pp. 195-200 (reprinted in Bell (2005), pp. 14-21).
3. Bell J.S. (1966), "On the problem of hidden variables in quantum mechanics", *Review of Modern Physics* **38**, pp. 447-452 ((reprinted in Bell (2005), pp. 1-13).
4. Bell J.S. (1971), "Introduction to the hidden-variable question", in B. D'Espagnat (ed.), *Foundations of Quantum Mechanics*, Academic Press, New York-London pp. 171-181 (reprinted in Bell (2005), pp. 29-39).
5. Bell J.S. (1981), "Bertlmann's socks and the nature of reality", *Journal de Physique* **42**, pp. 41-61 (reprinted in Bell (2005), pp. 139-158).
6. Bell J.S. (2005), *Speakable and Unspeakable in Quantum Mechanics*, Cambridge University Press, 2$^{nd}$ ed., Cambridge.
7. Butterfield J., Fleming G.N., Ghirardi G.C., Grassi R., "Parameter dependence and outcome dependence in dynamical models for state vector reduction", *Foundations of Physics* **23**, pp. 341-364.

8. Clauser J.F., Horne M.A., Shimony A., Holt R.A. (1969), "Proposed experiment to test local hidden-variable theories", *Physical Review Letters* **23**, pp. 880-884.

9. Clauser J.F., Horne M.A. (1974), "Experimental consequences of objective local theories", *Physical Review* **D10**, pp. 526-535.

10. Clauser J.F., Shimony A. (1978), "Bell's theorem: experimental tests and implications", *Reports on Progress in Physics* **41**, pp. 1881-1927.

11. Daumer M., Dürr D., Goldstein S., Maudlin T., Tumulka R., Zanghì N. (2006), "The message of the quantum?", in A. Bassi, D. Dürr, T. Weber and N. Zanghì (eds.), *Quantum Mechanics: Are there Quantum Jumps? On the Present Status of Quantum Mechanics*, AIP Conference Proceedings 844, American Institute of Physics, pp. 129-132.

12. D'Espagnat B. (1995), *Veiled Reality. An Analysis of Present-Day Quantum Mechanical Concepts*, Addison-Wesley, Mass.

13. Dürr D., Goldstein S., Zanghì N. (1996), "Bohmian mechanics as the foundations of quantum mechanics", in J.T. Cushing, A. Fine, S. Goldstein (eds.), *Bohmian Mechanics and Quantum Theory: an Appraisal*, Kluwer, Dordrecht, pp. 21-44.

14. Dürr D., Goldstein S., Zanghì N. (2004), "Quantum equilibrium and the role of operators as observables in quantum theory", *Journal of Statistical Physics* **116**, pp. 959-1055.

15. Fuchs C.A., Peres A. (2002), "Quantum Theory Needs No «Interpretation»", *Physics Today*, 53, pp. 70-71.

16. Giuntini R., Laudisa F. (2001), "The impossible causality: the no hidden variables theorem of John von Neumann", in M. Rédei, M. Stoeltzner (eds.), *John von Neumann and the Foundations of Quantum Physics*, Kluwer, Dordrecht, pp. 173-188.

17. Goldstein S. (1996), "Review essay: Bohmian mechanics and the quantum revolution", *Synthese* **107**, pp. 145-165.

18. Goldstein S. (2001), "Bohmian Mechanics", in E.N. Zalta (ed.), *Stanford Encyclopedia of Philosophy*, http://plato.stanford.edu/entries/qm-bohm

19. Gröblacher S., Paterek T., Kaltenbaek R., Brukner C., Žukowski M., Aspelmeyer M., Zeilinger A., (2007), "An experimental test of non-local realism", *Nature* **446**, pp. 871-875.

20. Hagar A. (2007), "Experimental metaphysics (take two)", *Studies in History and Philosophy of Modern Physics* **38**, pp. 906-919.

21. Isham C. (1995), *Lectures on Quantum Theory*, Imperial College Press, London.

22. Jarrett J. (1984), "On the physical significance of the locality conditions in the Bell arguments", *Nous* **18**, pp. 569-589.

23. Laudisa F. (1997), "Contextualism and nonlocality in the algebra of EPR observables", *Philosophy of Science* **64**, pp. 478-496.
24. Leggett A. (2003), "Nonlocal hidden-variable theories and quantum mechanics: an incompatibility theorem", *Foundations of Physics* **33**, pp. 1469-1493.
25. Maudlin T. (1996), "Space-time in the quantum world", in J. Cushing, A. Fine, S. Goldstein (eds.), *Bohmian Mechanics and Quantum Theory: an Appraisal*, Kluwer, Dordrecht, pp. 285-307.
26. Maudlin T. ($2002^2$), *Quantum Non-Locality and Relativity*, Blackwell, Oxford.
27. Nielsen M.N., Chuang I.L. (2000), *Quantum Computation and Information*, Cambridge University Press, Cambridge.
28. Norsen T. (2007), "Against 'realism' ", *Foundations of Physics* **37**, pp. 311-340.
29. Peres A., Terno D. (2004), "Quantum Information and Relativity Theory", *Review of Modern Physics* **76**, pp. 93-123.
30. Schrödinger E. (1935) "Die Gegenwärtige Situation in der Quantenmechanik", *Naturwissenschaften* **23**, pp. 807-812, 823-828, 844-849 (Engl. translation in Wheeler, Zurek 1983, pp. 152-167).
31. Shimony A. (1984), "Controllable and uncontrollable non-locality", in S. Kamefuchi et al. (eds.) *Foundations of Quantum Mechanics in Light of the New Technology*, he Physical Society of Japan, Tokyo (reprinted in A. Shimony, *Search for a Naturalistic Worldview*, Cambridge University Press, Cambridge 1993, vol. II, pp. 130-139.
32. Smerlak M., Rovelli C. (2007), "Relational EPR", *Foundations of Physics* **37**, pp. 427-445.
33. Squires E. (1986), *The Mystery of the Quantum World*, Adam Hilgher, Bristol.
34. Wheeler J.A., Zurek W.H. (eds.) (1983), *Quantum Theory and Measurement*, Princeton University Press, Princeton.
35. Zeilinger A. (2005), "The message of the quantum", *Nature* **438**, p. 743.
36. Žukowski M. (2005), "On the Paradoxical Book of Bell", *Studies in History and Philosophy of Modern Physics* 36, pp. 566-575.